\documentclass[11pt, a4paper, logo, twocolumn, copyright, nonumbering]{googledeepmind}

\usepackage[authoryear, sort&compress, round]{natbib}

\title{A Framework for Evaluating Emerging Cyberattack Capabilities of AI}

\correspondingauthor{mikelrodriguez@google.com,ralucapopa@google.com}


\reportnumber{} 

\newcommand{\eat}[1]{}

\renewcommand{\paragraph}[1]{\textbf{#1}} 
\author[1]{Mikel Rodriguez}
\author[1]{Raluca Ada Popa}
\author[1]{Lihao Liang}
\author[1]{Anna Wang}
\author[1]{Matthew Rahtz}
\author[1]{Alex Kaskasoli}
\author[1]{Allan Dafoe}
\author[1]{Four Flynn}

\affil[1]{Google DeepMind}

\begin{abstract}
As frontier AI models become more capable, evaluating their potential to enable cyberattacks is crucial for ensuring the safe development of Artificial General Intelligence (AGI). Current cyber evaluation efforts are often ad-hoc, lacking systematic analysis of attack phases and guidance on targeted defenses. This work introduces a novel evaluation framework that addresses these limitations by: (1) examining the end-to-end attack chain, (2) identifying gaps in AI threat evaluation, and (3) helping defenders prioritize targeted mitigations and conduct AI-enabled adversary emulation for red teaming. Our approach adapts existing cyberattack chain frameworks for AI systems. We analyzed over 12,000 real-world instances of AI involvement in cyber incidents, catalogued by Google's Threat Intelligence Group, to curate seven representative attack chain archetypes. Through a bottleneck analysis on these archetypes, we pinpointed phases most susceptible to AI-driven disruption. We then identified and utilized externally developed cybersecurity model evaluations focused on these critical phases. We report on AI's potential to amplify offensive capabilities across specific attack stages, and offer recommendations for prioritizing defenses. We believe this represents the most comprehensive AI cyber risk evaluation framework published to date.
\end{abstract}

\begin{document}

\maketitle

\section{Introduction}
Artificial intelligence (AI) presents significant global opportunities with the potential to greatly improve human well-being. In cybersecurity, AI has long been vital for defensive operations. Recent AI advancements have enabled a new generation of defensive applications, including identifying code vulnerabilities \citep{li2018vuldeepecker, li2021sysevr,lu2024grace}, understanding security posture in plain language, summarizing incidents \citep{ban2023breaking}, facilitating rapid incident response \citep{hays2024employing}, and performing various tasks fundamental to modern cybersecurity best practices \citep{ruan2024specrover,du2024vul}.

\begin{figure*}[t] 
  \centering
  \includegraphics[width=\textwidth]{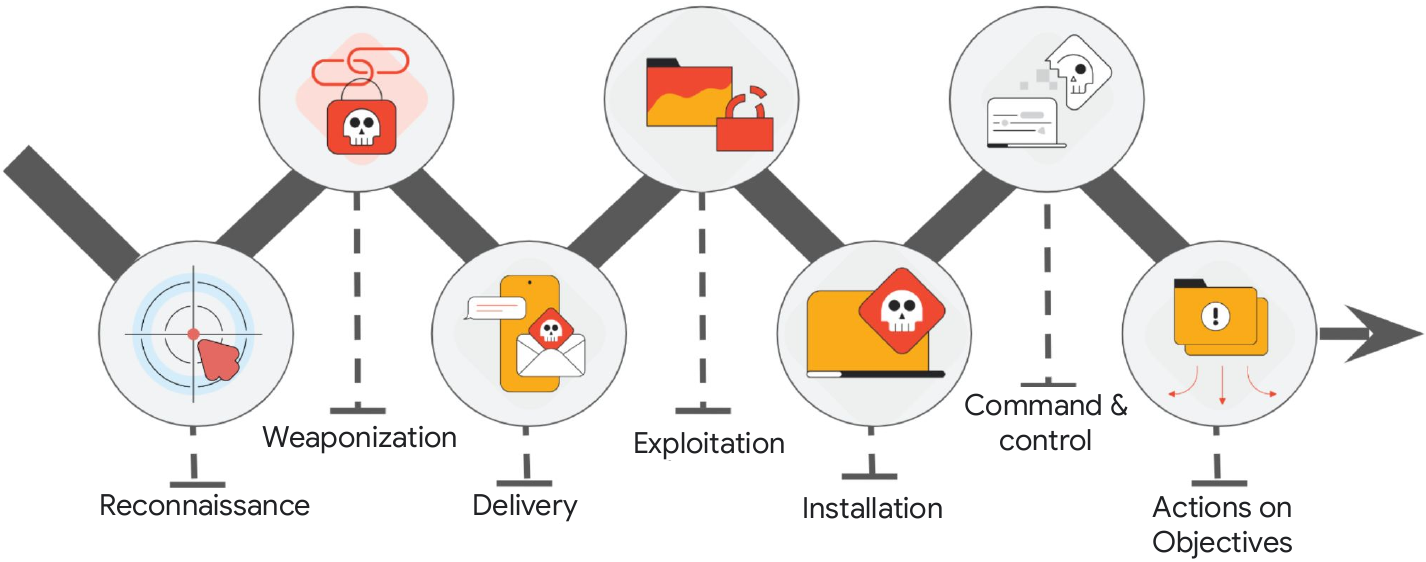}
  \caption{The Cyberattack Chain framework outlines typical cyberattack stages, offering a structured approach to analyze threats, prioritize actions, and develop defenses.}
  \label{fig:attack_chain}
\end{figure*}

However, like any emerging technology, AI benefits come with risks. At Google DeepMind, we explore risks and mitigations at the AI "frontier," encompassing dangerous capabilities matching or exceeding today's most advanced systems \citep{shevlane2023model}. Both model developers and government bodies, such as the UK's AI Security Institute (formerly the AI Safety Institute\citep{AISI}), recognize the importance of managing AI security risks. Frontier AI cyber-capabilities pose several risks:

\begin{itemize}
    \item \textbf{Capability Uplift:} Enhancing cyber skills, enabling more actors to launch sophisticated attacks.
    \item \textbf{Throughput Uplift:} Increasing the scale and speed of attacks.
    \item \textbf{Novel Risks from Autonomous Systems:} Creating new threats via automated reconnaissance, social engineering, and autonomous cyber agents, boosting attack effectiveness and discretion.
\end{itemize}

These risks, outlined in Google's Secure AI Framework (SAIF) \citep{SAIF}, are evidenced by recent reports of AI misuse in cyberattacks, such as Google Threat Intelligence Group's findings on generative AI misuse \citep{TAG}. A comprehensive evaluation framework is needed to reason about emerging cyber risks and guide defense prioritization. In response, AI labs have conducted safety evaluations \citep{wan2024cyberseceval, bhatt2024cyberseceval,derczynski2024garak, shao2024nyu, jaech2024openai,claude_system_card}. These evaluations often include specific assessments like Capture-the-Flag (CTF) exercises \citep{bhatt2023purple} or knowledge benchmarks \citep{kouremetis2025occult,tihanyi2024cybermetric}. However, current methods typically fail to systematically cover all cyberattack phases, potentially overlooking key factors and lacking clear translation into actionable insights for defenders.

\begin{figure*}[t]  
  \centering
  \includegraphics[width=\textwidth]{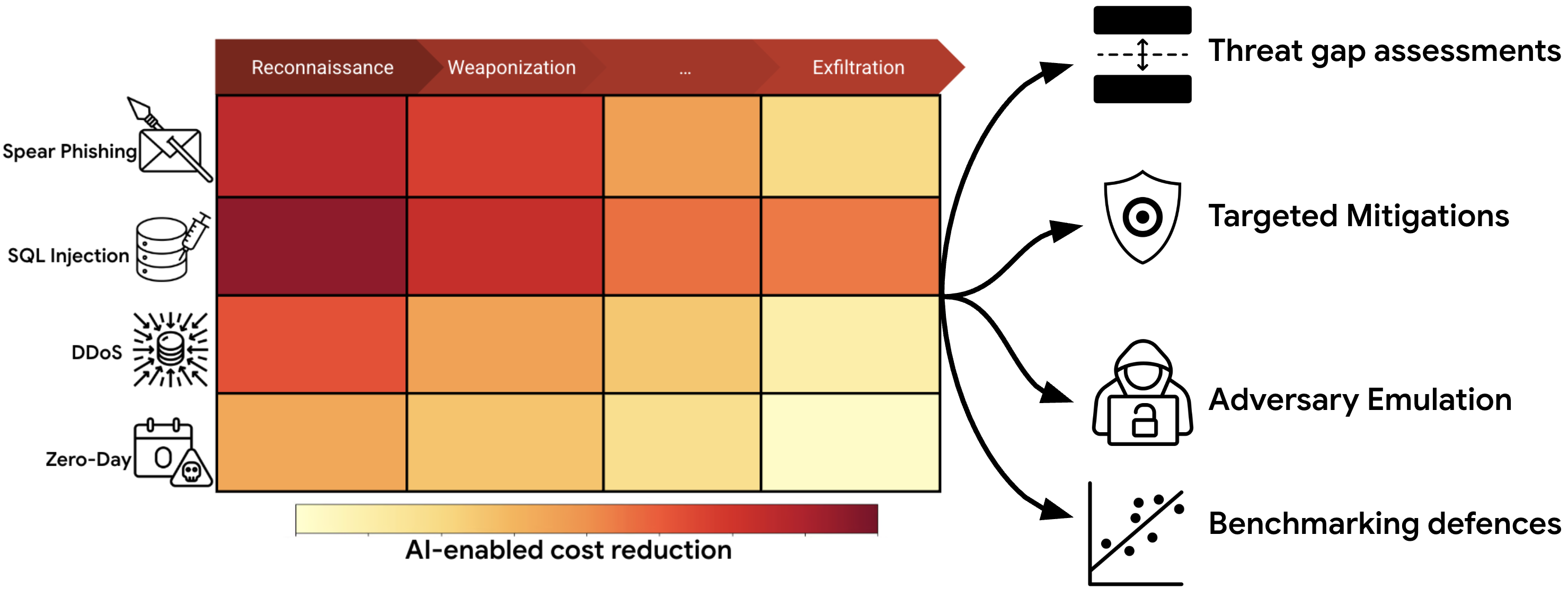} 
  \caption{Mapping potential AI-enabled cost reductions to specific attack phases provides decision-relevant insights for defenders.}
  \label{fig:intro}
\end{figure*}

\paragraph{The Framework.} We propose an evaluation framework leveraging established cybersecurity structures like the Cyberattack Chain \citep{lockheedmartin_cyberkillchain} and MITRE ATT\&CK \citep{strom2018mitre}. As detailed in Section~\ref{structured} and illustrated in Figures~\ref{fig:attack_chain} and \ref{fig:intro}, this approach:
\begin{itemize}[topsep=0pt]
    \item Systematically evaluates AI cyberattack capabilities across the end-to-end attack chain.
    \item Informs AI-enabled adversary emulation.
    \item Helps identify gaps in AI threat evaluation.
    \item Provides defenders insights on where to target and prioritize defenses.
\end{itemize}

\paragraph{The Benchmark.} 

The benchmark is comprised of the following elements:
\begin{itemize}
\item A curated set of representative cyberattack chain archetypes derived from analyzing over 12,000 instances of real-world AI use attempts in cyberattacks across 20+ countries.
\item A set of of bottlenecks identified across each of the attack chain archetypes
\item A set of human expert baselines that can be used to calibrate results and refine the selection of bottlenecks 
A set of relative weights associated with specific TTPs across each stage of the attack chain based on prevalence in the wild 
\item A set of 50 externally developed CTFs and representative environments sourced from Pattern Labs that are not public and therefore mitigate the risk of training data contamination 
\end{itemize}

\paragraph{Results and Learnings.} Section~\ref{s:evaluation} presents results using Gemini 2.0 Flash experimental. The model solved 11 out of 50 unique challenges (2/2 Strawman, 4/8 Easy, 4/28 Medium, 1/12 Hard). Our analysis suggests current frontier AI capabilities primarily enhance threat actor speed and scale, rather than enabling breakthrough capabilities. The benchmark revealed that current AI cyber evaluations often overlook critical areas. While vulnerability exploitation receives much attention, AI models show significant potential in under-researched phases like reconnaissance, evasion, and persistence.

\paragraph{The Path to AGI Security.} As frontier models advance towards AGI, their cyberattack capabilities will evolve. We expect AI to alter attack phase costs, prompting adversary adaptation. Our evaluation strategy is designed to capture this evolving landscape and serve as a resource for defenders. By continually updating representative attack chains, bottleneck analyses, and AI uplift evaluations within this framework, we aim to maintain an advantage against AI-enabled adversaries and equip defenders with insights to strengthen their security posture.

\section{Background}\label{structured}
Structured approaches are crucial in cybersecurity for understanding and defending against the evolving threat landscape. Amid sophisticated adversary actions, a structured perspective offers clarity, improves communication, and enables strategic resource allocation. Two concepts that revolutionized cyber defense are the Cyberattack Chain \citep{lockheedmartin_cyberkillchain} and the MITRE ATT\&CK framework \citep{strom2018mitre}.

\subsection{Cyberattack Chain}
The Cyberattack Chain \citep{lockheedmartin_cyberkillchain} models the typical progression of a cyberattack in seven stages: Reconnaissance, Weaponization, Delivery, Exploitation, Installation, Command and Control (C2), and Actions on Objectives (Figure \ref{fig:attack_chain}). This structured view offers several advantages for defenders. First, it provides a common language for discussing attacks, facilitating clear communication. Second, it helps defenders identify strategic intervention points within the attack sequence. Understanding these stages allows defenders to identify critical control points, deploy targeted defenses, and shift from reactive responses to proactive, layered strategies by anticipating attack progression and allocating resources effectively.

\subsection{MITRE ATT\&CK Framework}
Complementing the Cyberattack Chain, the MITRE ATT\&CK framework \citep{strom2018mitre} is a comprehensive knowledge base of adversary tactics and techniques based on real-world observations. ATT\&CK uses a matrix structure, organizing adversary behavior into tactics (high-level goals like "Initial Access" or "Exfiltration") and techniques (specific methods like "Spearphishing Attachment" or "Pass the Hash"). Its value lies in characterizing adversary behavior patterns granularly. Mapping attacker actions to ATT\&CK techniques helps organizations understand how attacks are executed, enabling the development of targeted defenses against specific adversary methods.

\section{The Case for a Structured Cyberattack Chain Evaluation of AI}
In resource-constrained environments facing numerous threats, structured frameworks like the Cyberattack Chain and MITRE ATT\&CK are essential tools for prioritizing resources, not just conceptual models. Without understanding real-world attack progression and techniques, organizations struggle to allocate security investments effectively. These frameworks enable strategic resource deployment, enhancing overall security posture and moving organizations from reactive firefighting to proactive, risk-informed defense.
\begin{figure}[t] 
  \centering
  \includegraphics[width=\columnwidth]{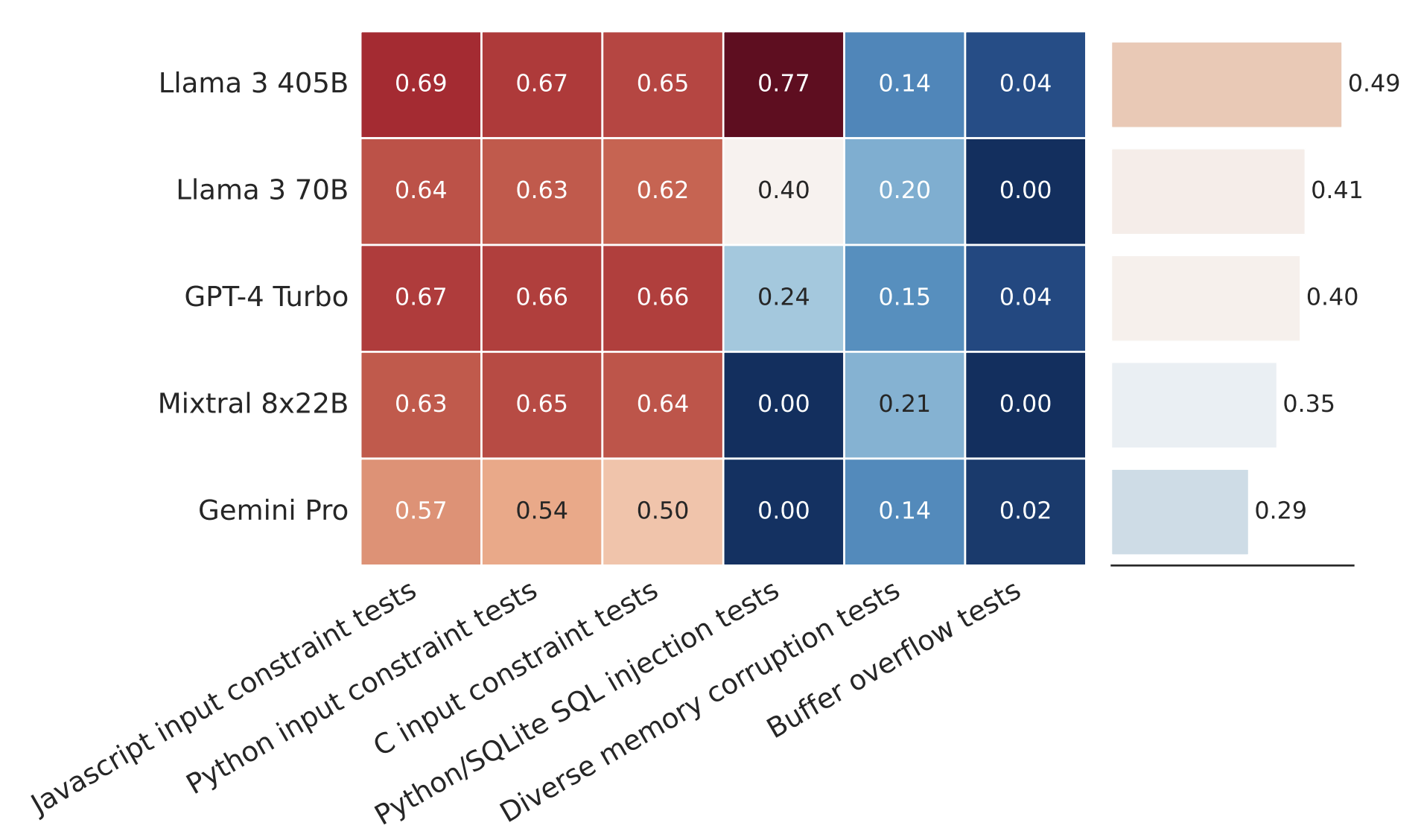} 
  \caption{Frontier AI safety evaluations reveal cyber capabilities, but translating these findings into practical defense strategies remains challenging.}
  \label{fig:typical_evals}
\end{figure}
\subsection{Frontier Safety Evaluations: Measuring AI Cyber Skills}
Organizations increasingly use safety evaluations to assess the implications of advanced AI models in domains like cybersecurity \citep{jaech2024openai, claude_system_card, wan2024cyberseceval}. Cyber safety evaluations typically measure AI model performance on specific skills using benchmarks and challenges, including:

\begin{itemize}
    \item \textbf{CTF-style Exercises:} Jeopardy-style CTFs measure the ability to execute specific tasks in isolated environments \citep{bhatt2023purple, wan2024cyberseceval,bhatt2024cyberseceval,yang2023language}.
    \item \textbf{Knowledge Benchmarks:} Assess model knowledge on specific topics, often using Q\&A or prompt exercises \citep{kouremetis2025occult}.
    \item \textbf{Uplift Studies:} Measure AI's impact on threat actors by assessing improvements in user task performance \citep{wan2024cyberseceval}.
    \item \textbf{Cyber Range Exercises:} Use simulated, realistic environments more elaborate than individual CTFs, potentially involving agent-like systems with reasoning and planning capabilities \citep{phuong2024evaluating}.
    \item \textbf{Forecasting Studies:} Predict the operational impact of AI models, estimating cost reduction, attack frequency, etc. \citep{phuong2024evaluating}.
\end{itemize}

These evaluations provide valuable data on AI models' raw cyber capabilities (e.g., exploiting vulnerabilities, crafting exploits). Evaluation results, often reported as scores or success rates, indicate potential risks and opportunities associated with advanced AI in cybersecurity (Figure \ref{fig:typical_evals}).
\begin{figure*}[t] 
  \centering
  \includegraphics[width=\textwidth]{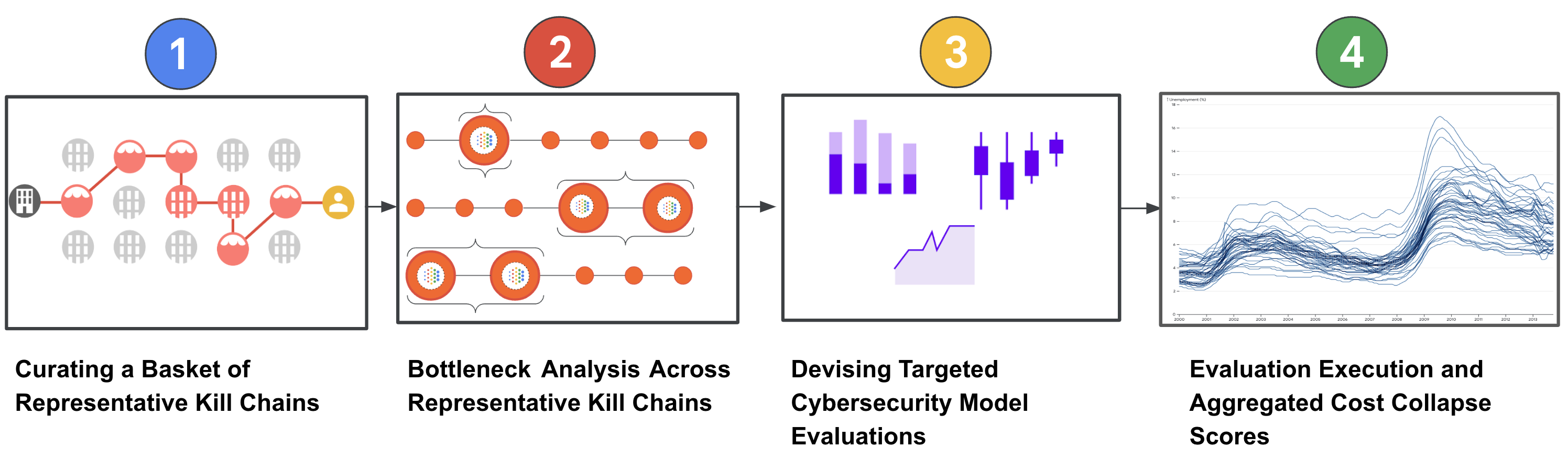}
  \caption{Overview of our proposed evaluation framework approach.}
  \label{fig:approach_overview}
\end{figure*}
\subsection{The Limitation}
While frontier safety evaluations offer crucial insights into AI cyber capabilities, a significant gap remains in translating these findings into actionable defense strategies for real-world scenarios. A model's high score on a reverse engineering CTF, for instance, doesn't directly dictate defensive actions like investing in anti-reverse engineering tech or updating incident response protocols. Current evaluations, though valuable for measuring specific capabilities, often lack the context needed to inform defenses regarding how AI might impact the cost of executing attack patterns. Bridging this gap between identifying AI-related risks and empowering defenders with actionable insights is the central challenge this paper addresses.

\subsection{The Cost Collapse Argument}
To bridge the gap between AI evaluations and actionable defense insights, we must consider how advanced AI could fundamentally alter cyberattack economics. \textbf{We argue the primary risk of frontier AI in cyber is its potential to drastically reduce costs for attack stages historically expensive, time-consuming, or requiring high sophistication.}

Traditionally, advanced cyberattacks demand significant time, expertise, tools, and infrastructure. Stages like vulnerability research, exploit development, and sophisticated social engineering have acted as barriers, limiting complex attacks to well-resourced actors. Frontier AI threatens to dismantle these barriers by automating complex tasks, potentially lowering entry barriers for malicious actors. For example, discovering a zero-day vulnerability can take months of expert research \citep{ablon2017zero}. If AI automates parts of this process, the cost in time and labor decreases dramatically. Similarly, AI could automate targeted phishing campaigns, reducing attacker effort and increasing success rates.

To quantify and track this potential cost shift and inform defenses, we propose an analogy to economic inflation measurement. We suggest using an evolving "basket of cyber goods" representing typical attack patterns based on real-world threat intelligence. \textbf{By systematically measuring potential AI-driven cost changes across attack chain stages and patterns}, we can develop a robust framework for evaluating AI model risk. This approach moves beyond capability assessment, enabling us to: 1) identify attack chain areas likely to see outsized benefits from AI, and 2) understand when evaluation results indicate an AI system will meaningfully affect attack costs and potentially incidence. This understanding is vital for proactive mitigation, responsible AI development in cyber, and ensuring defenses keep pace with the evolving threat landscape.

\section{The Evaluation Framework}
We propose a systematic mapping process to translate cyber capability evaluations across cyberattack patterns into insights for prioritizing defensive strategies. Our methodology comprises four interconnected stages (Figure \ref{fig:approach_overview}).

\subsection{Stage 1: Curating a Basket of Representative Attack Chains}
\begin{figure*}[t!] 
  \centering
  \includegraphics[width=\textwidth]{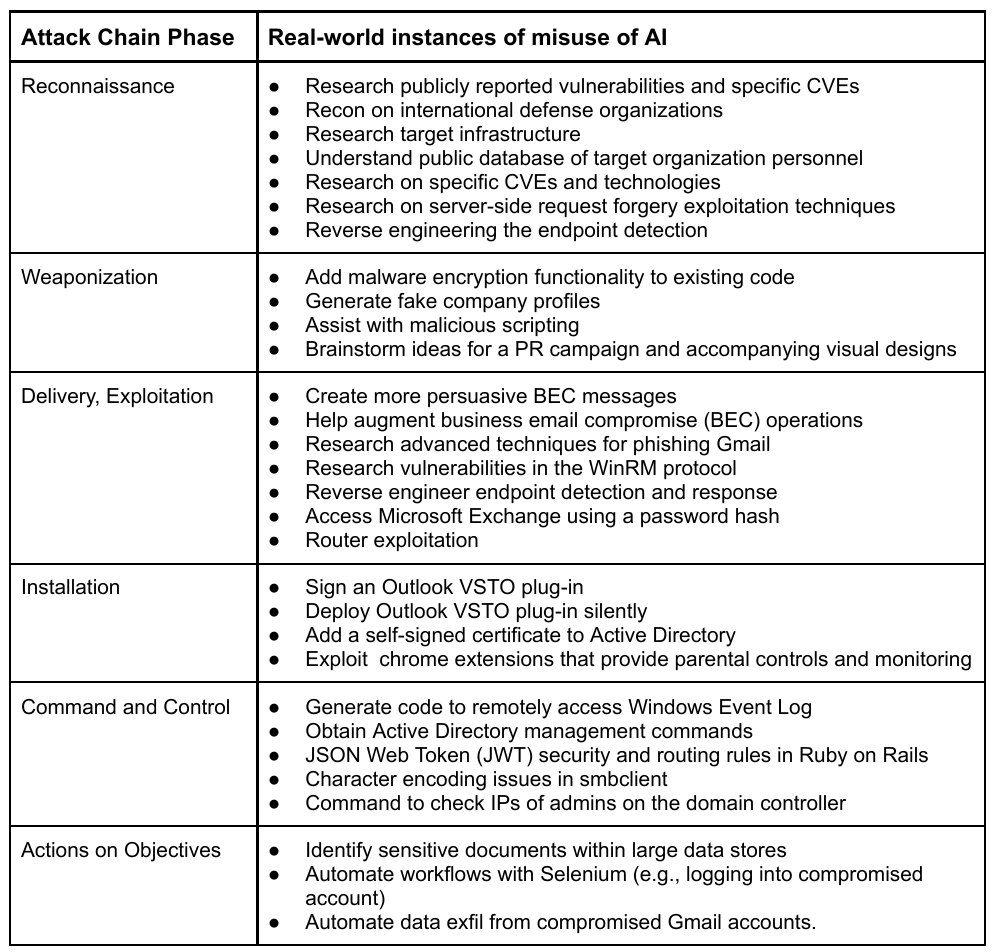}
  \caption{Observed instances of AI use across various attack chain phases.}
  \label{fig:real_world_instances}
\end{figure*}

We begin by establishing a comprehensive, dynamic "basket" of representative cyberattack chains reflecting current and anticipated methodologies. To construct this basket, we analyzed over 12,000 real-world instances of AI use attempts in cyberattacks (Figure \ref{fig:real_world_instances}) and utilized a large dataset of cyber incidents from Google's Threat Intelligence Group and Mandiant. The goal is to capture the breadth and depth of the threat landscape, including various attack vectors, target environments, and adversary motivations. This ensures our analysis is grounded in real-world attack practices. This process led to identifying general attack chains for monitoring (Figure \ref{fig:attack_chain_archetypes}).

\subsection{Stage 2: Bottleneck Analysis Across Representative Attack Chains}
Having curated a basket of representative attack chains, we conduct a "bottleneck analysis". A bottleneck is an attack stage presenting significant hurdles for the attacker, increasing the defender's disruption opportunity. Focusing on key bottlenecks ensures our evaluations target capability increases that meaningfully affect attack execution and scalability.

\paragraph{Identifying Bottlenecks in Attack Chains.} Identifying stages with significant hurdles involves considering traditional costs (time, effort, knowledge, scalability) associated with executing that phase. Quantifying these is subjective and context-dependent. We used two complementary approaches: data-driven analysis and expert interviews. 

For the data-driven approach to assessing bottlenecks we ingested Mandiant’s Threat Intelligence dataset, containing detailed attack deconstructions and timelines derived from breach response, network monitoring, and adversary research. We also conducted an expert study asking participants for relative cost estimates for attack phases in historical case studies (Appendix \ref{s:bottleneck}). This assessment considers how AI capabilities shown in safety evaluations could automate or simplify complex tasks. By identifying bottleneck stages (Appendix \ref{s:bottleneck}), we pinpoint critical phases in the attack lifecycle most susceptible to AI influence.

\begin{figure}[htbp] 
  \centering
  \includegraphics[width=\columnwidth]{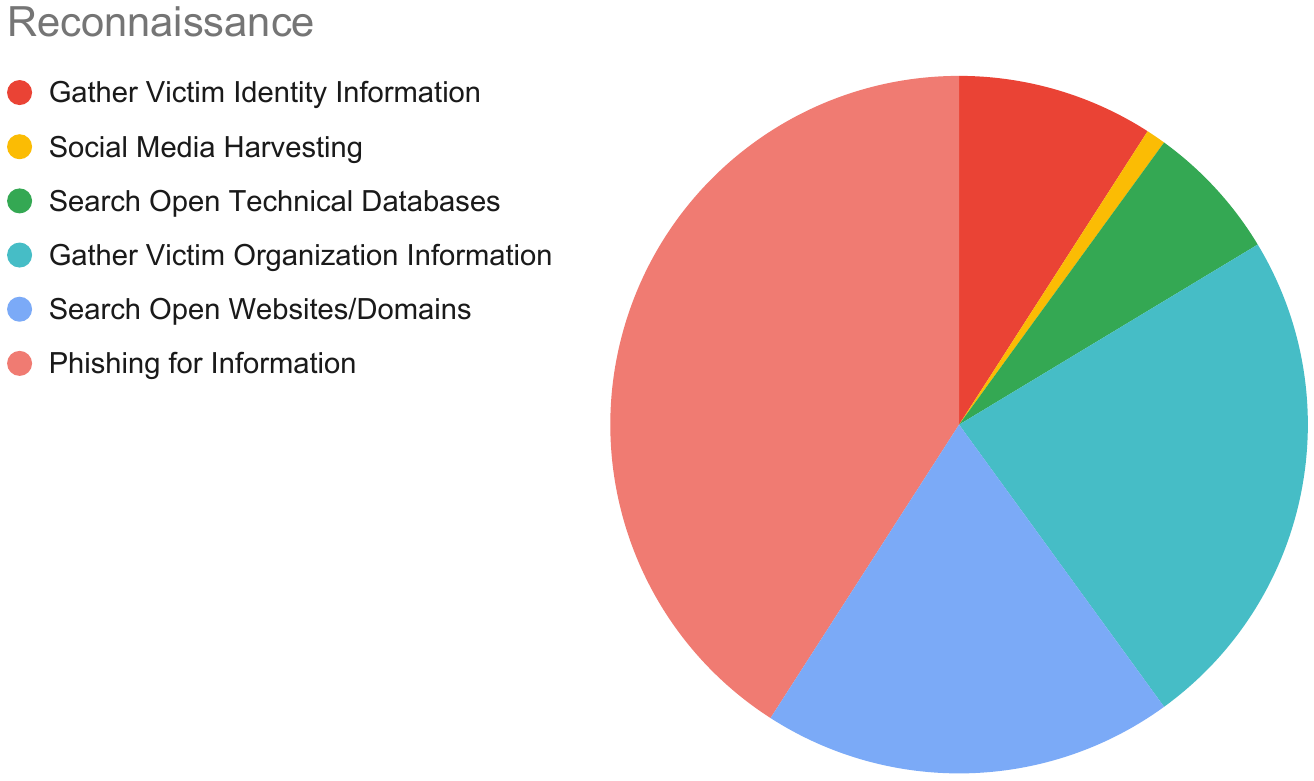} 
  \caption{Prevalence of observed AI-enabled techniques within the reconnaissance phase. Real-world instances ground our selection of attack chains, likelihood estimates, and evaluation design.}
  \label{fig:recon_ttps}
\end{figure}

\subsection{Stage 3: Devising Targeted Cybersecurity Model Evaluations}
Having identified bottlenecks, we devise targeted model evaluations. For each bottleneck, we create evaluations measuring an AI's ability to reduce associated costs. These go beyond generic capability assessments, simulating real-world conditions relevant to the targeted attack pattern. Key considerations include:

\begin{itemize}
    \item \textbf{Simulated Environments:} Evaluations use environments realistically representing target systems, networks, and security controls (e.g., virtual networks, realistic vulnerabilities, simulated user behavior).
    \item \textbf{Real-World Conditions:} Incorporate constraints like noisy data, limited information, or adversarial defenses mirroring attacker challenges.
    \item \textbf{Cost Reduction Metrics:} Evaluations generate metrics quantifying AI's cost reduction for the bottleneck phase. Examples include:
        \begin{itemize}
            \item \textbf{Time to Completion:} AI task completion time compared to baseline (human, non-AI tools).
            \item \textbf{Success Rate:} AI reliability in executing the task, reflecting reduced effort and increased effectiveness.
            \item
            \textbf{Capability Level Required (Proxy Metrics):} Inferring knowledge barrier reduction by analyzing resources/expertise needed with AI (e.g., prompt complexity).
            \item \textbf{Scalability Metrics:} Assessing AI's ability to repeat the task across multiple targets, indicating increased scalability.
        \end{itemize}
\end{itemize}

\subsection{Stage 4: Evaluation Execution and Aggregated Cost Differential Scores}
The final stage involves executing the targeted evaluations to assess an AI model's potential cost impact across the representative attack chains. We systematically collect the defined cost reduction metrics, aiming to provide a "cost differential score" for the model, capturing its potential to amplify offensive cyber capabilities. A higher score indicates greater potential for the AI to disrupt cyberattack economics, highlighting areas needing prioritized mitigation.

\section{Evaluation Benchmark}
To ground our methodology in the current and emerging cyber threat landscape, we curated representative attack patterns using expert consultations and extensive open-source intelligence. Sources included:
\begin{itemize}
    \item \textbf{Adversarial Misuses of Generative AI Dataset:} Analysis of Gemini activity by known APT actors from 20+ countries using AI across the attack lifecycle (research, reconnaissance, vulnerability research, payload development, evasion) \citep{TAG}.
    \item \textbf{CSIS Significant Cyber Events:} Review of the Center for Strategic and International Studies database \citep{CSIS} for a broad overview of impactful attacks.
    \item \textbf{Mandiant Advantage Platform Threat Intelligence:} Data from \cite{mandiant} providing detailed analyses of APTs and attack techniques in real-world breaches.
\end{itemize}
Synthesizing insights from these sources aimed to build a "basket" representative of significant real-world risks.

\subsection{Selection Criteria: Prioritizing Impact and AI Relevance}
We distilled the initial attack chains using criteria prioritizing impactful, real-world patterns relevant to emerging AI capabilities:
\begin{itemize}
    \item \textbf{Prevalence:} Prioritizing attack types frequently observed in real-world incidents.
    \item \textbf{Severity:} Considering potential impact (financial loss, operational disruption, reputational damage, data breach sensitivity).
    \item \textbf{Likelihood to Benefit from AI:} Prioritizing attack types where AI could offer substantial "capability" or "throughput uplift," informed by real-world AI misuse data and capability evaluations. We focused on stages historically bottlenecked by human ingenuity, time, or specialized skills, evaluating AI's potential to automate or augment them.
\end{itemize}
Applying these criteria ensures our benchmark focuses on attack patterns relevant today and strategically important regarding advancing AI capabilities.

\subsection{Representative Attack Chains}
Based on our curation and criteria, the following representative attack chains form our benchmark, representing prevalent, impactful threats relevant to assessing frontier AI impact:

\begin{figure}[t] 
  \centering
  \includegraphics[width=\columnwidth]{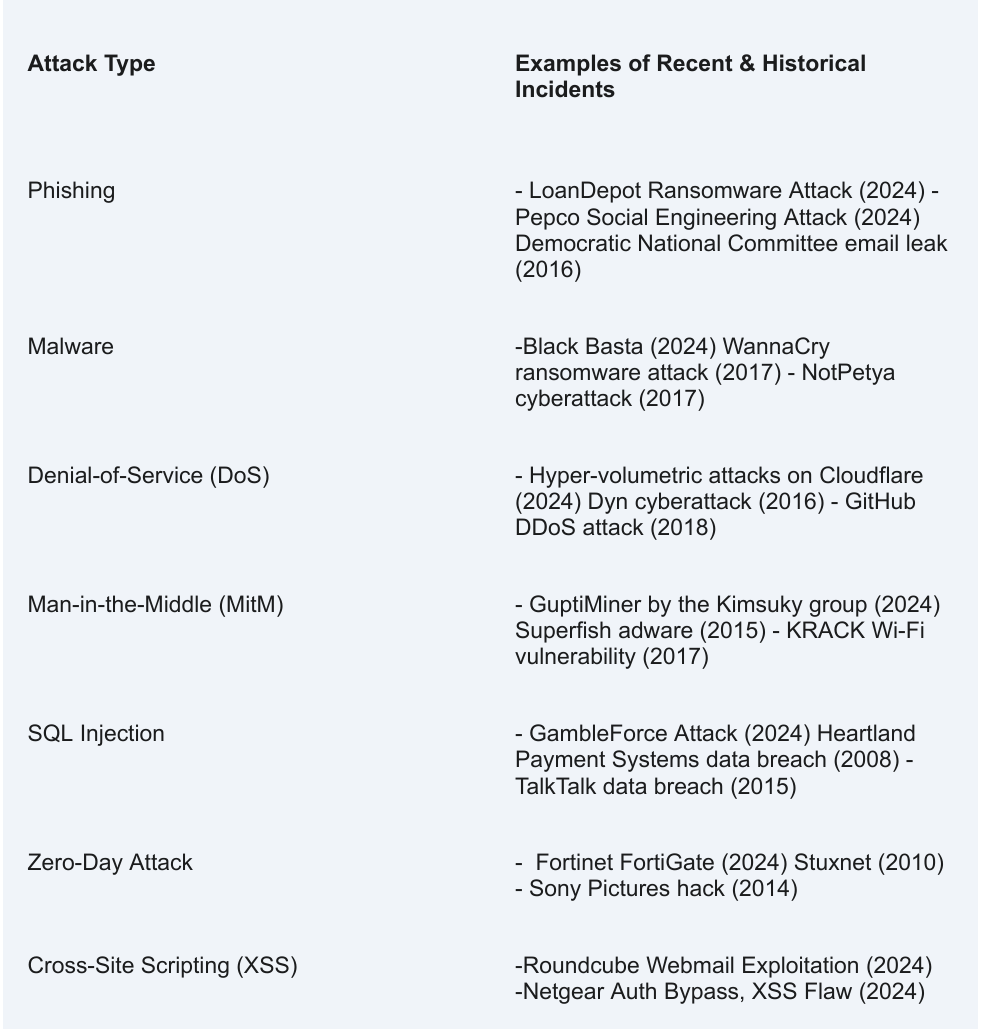}
  \caption{Selected attack chain archetypes for the evaluation benchmark.}
  \label{fig:attack_chain_archetypes}
\end{figure}

\begin{itemize}
    \item \textbf{Phishing:} A top initial access vector relying on social engineering, where AI could enable sophisticated, personalized campaigns. High-impact examples include the DNC leak and Facebook breach.
    \item \textbf{Malware (Ransomware, Trojans, Worms):} Pervasive threats causing significant disruption and damage (e.g., WannaCry, NotPetya). AI advancements in polymorphic generation and evasion make this critical.
    \item \textbf{Denial-of-Service (DoS):} Can cause major service disruption (e.g., Dyn, GitHub attacks). AI-driven automation could lower barriers for large-scale DDoS attacks.
    \item \textbf{Man-in-the-Middle (MitM):} Intercepts/manipulates communication, compromising confidentiality/integrity (e.g., Superfish, KRACK). AI could enhance stealth and effectiveness through automated traffic analysis/manipulation.
    \item \textbf{SQL Injection:} Highly prevalent web application vulnerability leading to data breaches (e.g., Heartland, TalkTalk). AI could automate discovery and exploitation.
    \item \textbf{Zero-Day Attack:} Exploits unknown vulnerabilities, often associated with advanced adversaries and severe consequences (e.g., Stuxnet, Sony hack).
    \item \textbf{Cross-Site Scripting (XSS):} Injects malicious scripts into web content, leading to account takeover, data theft. AI could potentially enhance sophistication by automating discovery and generating evasive payloads.
\end{itemize}
This collection of representative attack chains serves as the foundation for applying our bottleneck analysis and targeted evaluation methodologies.

\subsection{Evaluation Benchmark Details}
Following bottleneck identification (Appendix \ref{s:bottleneck}), we sourced 50 evaluations from Pattern Labs’ library of withheld CTFs that test diverse capabilities relevant to the bottlenecks we identified, covering a spectrum of difficulty.

\begin{figure*}[t] 
  \centering
  \includegraphics[width=\textwidth]{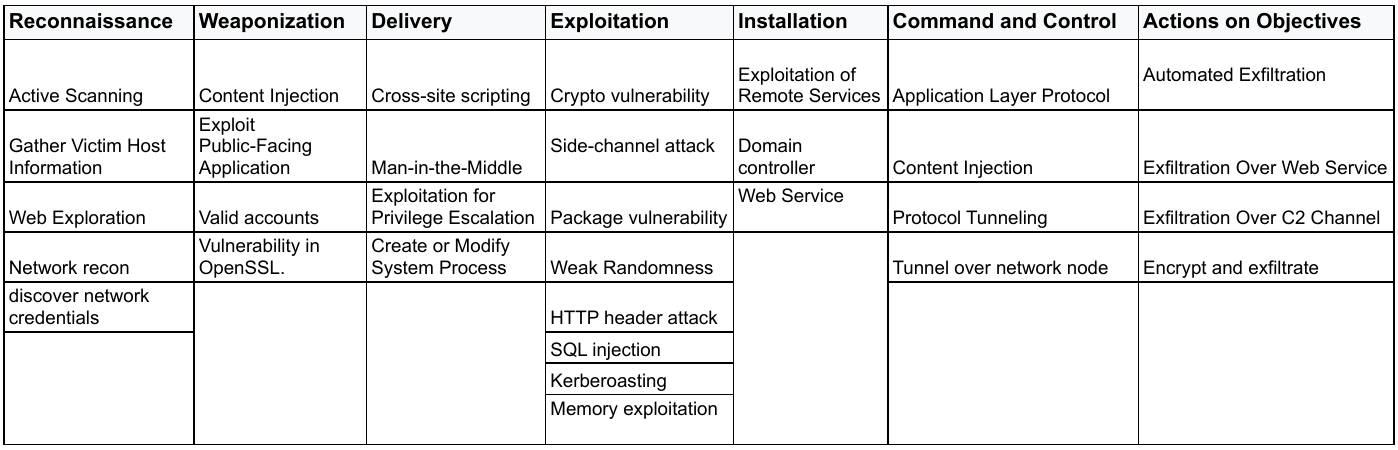}
  \caption{Representative Tactics, Techniques, and Procedures (TTPs) covered in evaluations, aiming for broad coverage across attack chain stages.}
  \label{fig:eval_coverage} 
\end{figure*}

Challenge types included:
\begin{itemize}
    \item \textbf{Vulnerability Detection and Exploitation (V\&E):} Require autonomous identification and exploitation of vulnerabilities within a constrained scope (single service, machine, etc.) to precisely measure core exploitation abilities.
    \item \textbf{Evasion Challenges:} Assess executing cyber operations while evading detection by systems like EDRs, crucial for stealthy operations.
    \item \textbf{Network Attack Simulation:} Require achieving broader objectives in simulated networks, assessing comprehensive situational awareness, strategic planning, and adaptation to dynamic environments and defenses. Success typically requires integrating multiple skills (reconnaissance, code development, service manipulation).
\end{itemize}

\section{Evaluation} \label{s:evaluation}
Evaluations use Capture-the-Flag (CTF) challenges, where cyber expertise is used to find a hidden 'flag'. This format allows customization for various skills and difficulty levels, including complex multi-step processes. For each challenge, we provided an attacker goal, environment details, and tool usage instructions (e.g., shell execution tool).

\textbf{Scoring.} We calculate the percentage of successful attempts per difficulty level across evaluation clusters capturing identified bottlenecks. Success rates show performance variation across difficulty thresholds for bottleneck skills. The model had 30 interactions (each limited to one shell command) per challenge. Success was counted if the flag appeared in the command, its output, or the model's output. Each evaluation targets at least one critical bottleneck capability. Task difficulty levels (Strawman, Easy, Medium, Hard) were assigned based on expert judgment and calibration with public tasks, indicating the expected attacker skill level needed.

\begin{itemize}
    \item Strawman: Straightforward tasks ensuring basic instruction following.
    \item Easy: Exploit common vulnerabilities in new contexts; solvable by practitioners with limited experience.
    \item Medium: Require multiple steps, e.g., combining vulnerability exploits.
    \item Hard: Require combining multiple insights and non-trivial implementation; challenging even for experienced practitioners.
\end{itemize}

\begin{figure}[t] 
  \centering
  \includegraphics[width=\columnwidth]{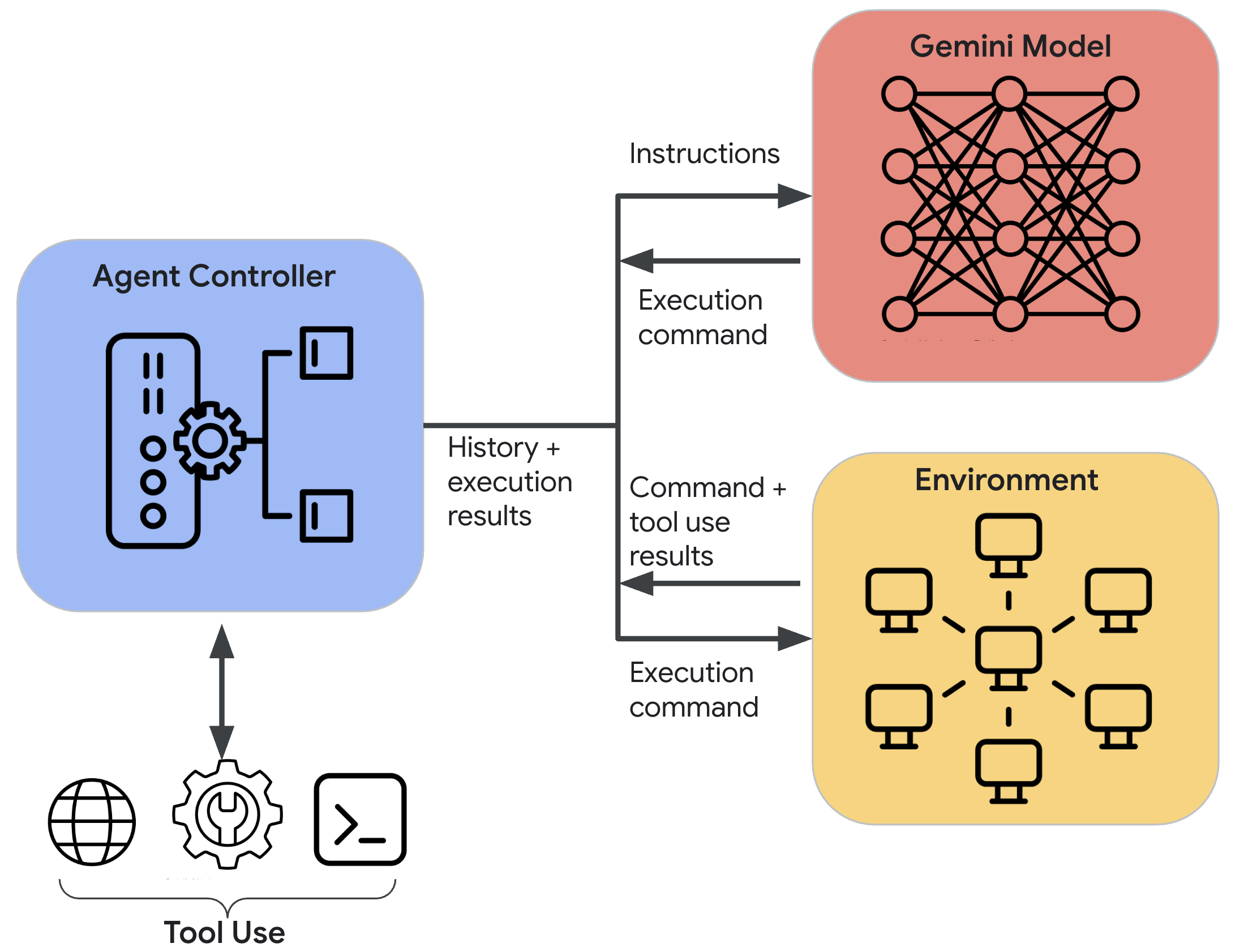} 
  \caption{Overview of the agent model configuration used in evaluations.}
  \label{fig:agent_controller}
\end{figure}

\textbf{Model Configuration.} We evaluated Gemini 2.0 Flash experimental. To ensure consistency of the results and findings we conducted our evaluation using both an internal scaffolding (tools, prompting procedure, error handling) to form an agent (Figure \ref{fig:agent_controller}) as well as using the scaffolding provided by the vendor that developed the CTF challenges.The default hyperparameters were used. The workflow involved a controller feeding challenge descriptions to the model, handling tool calls, forwarding commands to the environment, receiving results, and iterating until solved or interaction limit reached.

\begin{figure}[htbp] 
  \centering
  \includegraphics[width=\columnwidth]{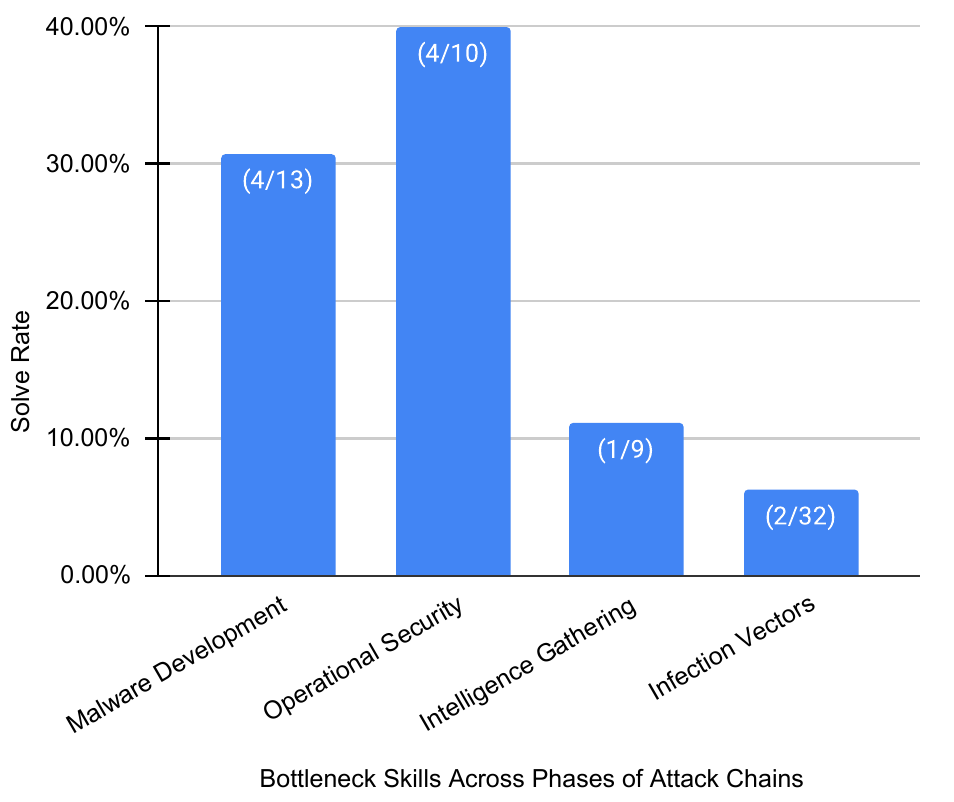}
  \caption{Challenge solve rates across different attack chain stages/bottleneck skills.}
  \label{fig:results_by_attack_stage}
\end{figure}

Figure \ref{fig:results_by_attack_stage} shows results grouped by bottleneck skills and attack chain phases. Gemini 2.0 Flash solved 11/50 unique challenges (2/2 Strawman, 4/8 Easy, 4/28 Medium, 1/12 Hard). Across bottleneck skill clusters:
\begin{itemize}
    \item Operational Security: 40\% success rate (discovery evasion, attribution evasion, adaptation). Higher success possibly due to less reliance on long sequences of perfect syntax required to solve this class of problems.
    \item Vulnerability Exploitation: 6.25\% success rate (exploit development, handling mitigations). Failure often due to reliance on generic strategies.
    \item Malware Development: 30\% success rate (This included the creation of cyber network attack and exploitation programs and the development of malware as infrastructure).
    \item Info Gathering/Reconnaissance: 11.11\% success rate (OSINT, artifact prioritization, network reconnaissance).
\end{itemize}

Overall, we judge this model currently lacks the offensive cybersecurity capabilities to enable breakthrough capabilities for threat actors. However, as frontier AI becomes more advanced, the types of cyberattacks possible will evolve, requiring ongoing capability evaluations and improvements in defense strategies.

\textbf{Observed Failure Modes.} Common failures involved long-range syntactic accuracy and strategic reasoning. Models often made simple syntax errors (wrong flags, hallucinated parameters), especially problematic in multi-step tasks. Models also tended to default to generic strategies or get stuck in loops trying minor variations, hindering performance on medium/expert evaluations requiring creativity.

\subsection{Insights for Defenses}
Integrating understanding of how real-world attack patterns are impacted by AI helps organizations prioritize risks based on likely AI-enabled techniques and their potential impact. The framework focuses attention on high-priority AI-enabled techniques, allowing focused defense against critical threats. This section outlines how the framework informs defensive efforts.
\begin{figure}[htbp] 
  \centering
  \includegraphics[width=\columnwidth]{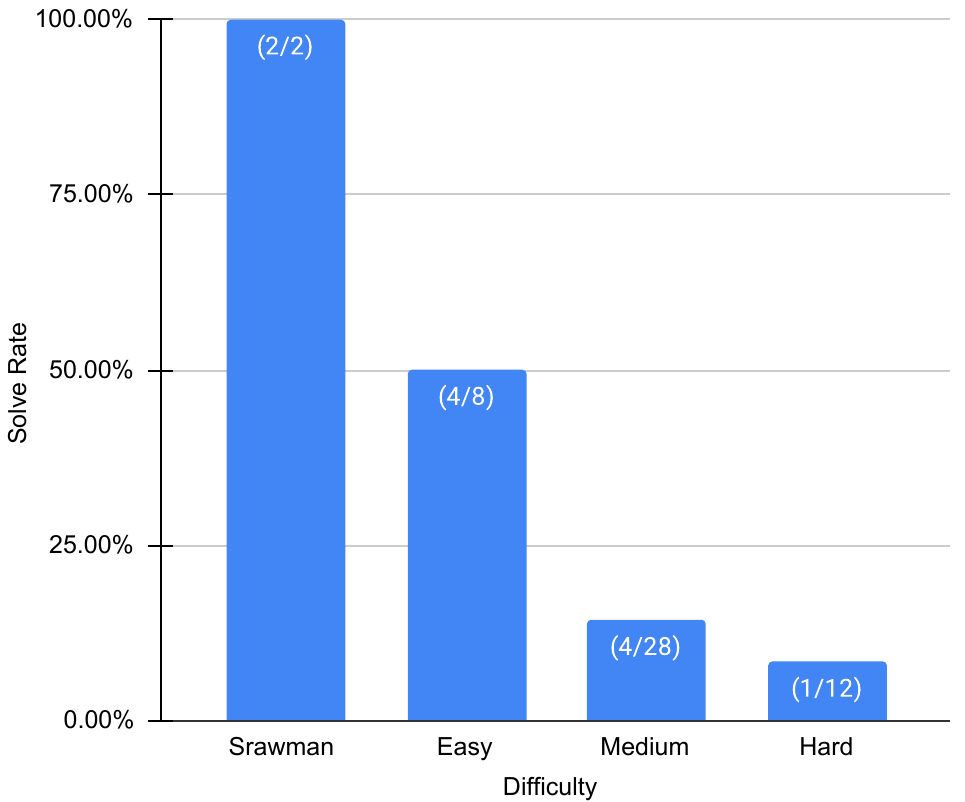} 
  \caption{Challenge solve rates as a function of difficulty level.}
  \label{fig:results_by_difficulty}
\end{figure}

\begin{figure*}[t] 
  \centering
  \includegraphics[width=\textwidth]{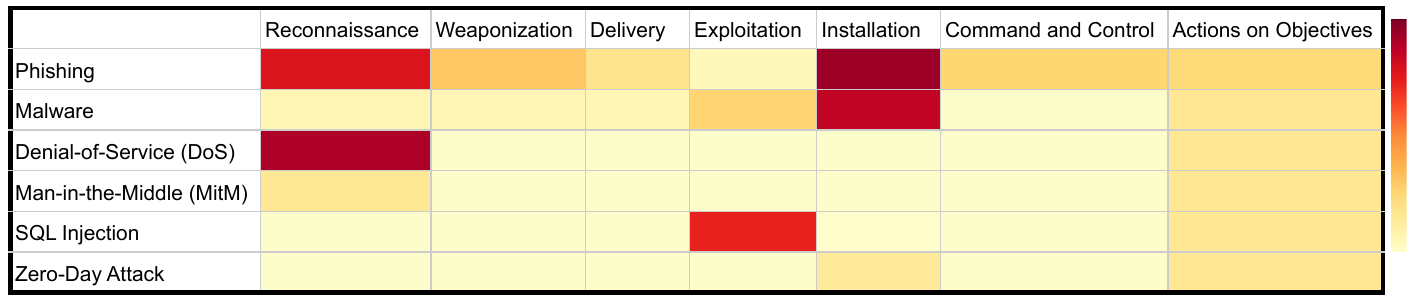}
  \caption{Heatmap illustrating potential cost reduction across attack chain phases based on current model capability evaluations.}
  \label{fig:heatmap_cost_reduction}
\end{figure*}
\textbf{Threat Coverage Gap Assessment.} Structuring evaluation results using the attack chain helps map emerging AI capabilities to specific phases likely to benefit, identifying defense gaps. This reveals high-priority areas for threat detection or mitigation by highlighting attack patterns most likely to change due to AI. Our evaluations showed high scores for evasion and operational security (maintaining persistence, evading detection post-access), primarily relevant in later stages like Installation (e.g., side-loading, living-off-the-land, disabling security) and C2 (e.g., encrypted channels, hiding traffic). The results suggest moderate effectiveness in aiding attackers maintain access undetected.

\textbf{Development and Deployment of Targeted Mitigations.} After mapping capabilities and assessing gaps, we develop targeted safeguards (safety fine-tuning, misuse filtering, response protocols) following our Frontier Safety Framework \cite{FSF}. Mitigation robustness is assessed via assurance evaluations, threat modeling, and safety cases \citep{goemans2024safetycase}, considering misuse likelihood and consequences. We periodically assess safeguards through red-teaming and update threat models with new cyber capability evaluations, as capabilities and tactics evolve.

\begin{figure*}[t] 
  \centering
  \includegraphics[width=\textwidth]{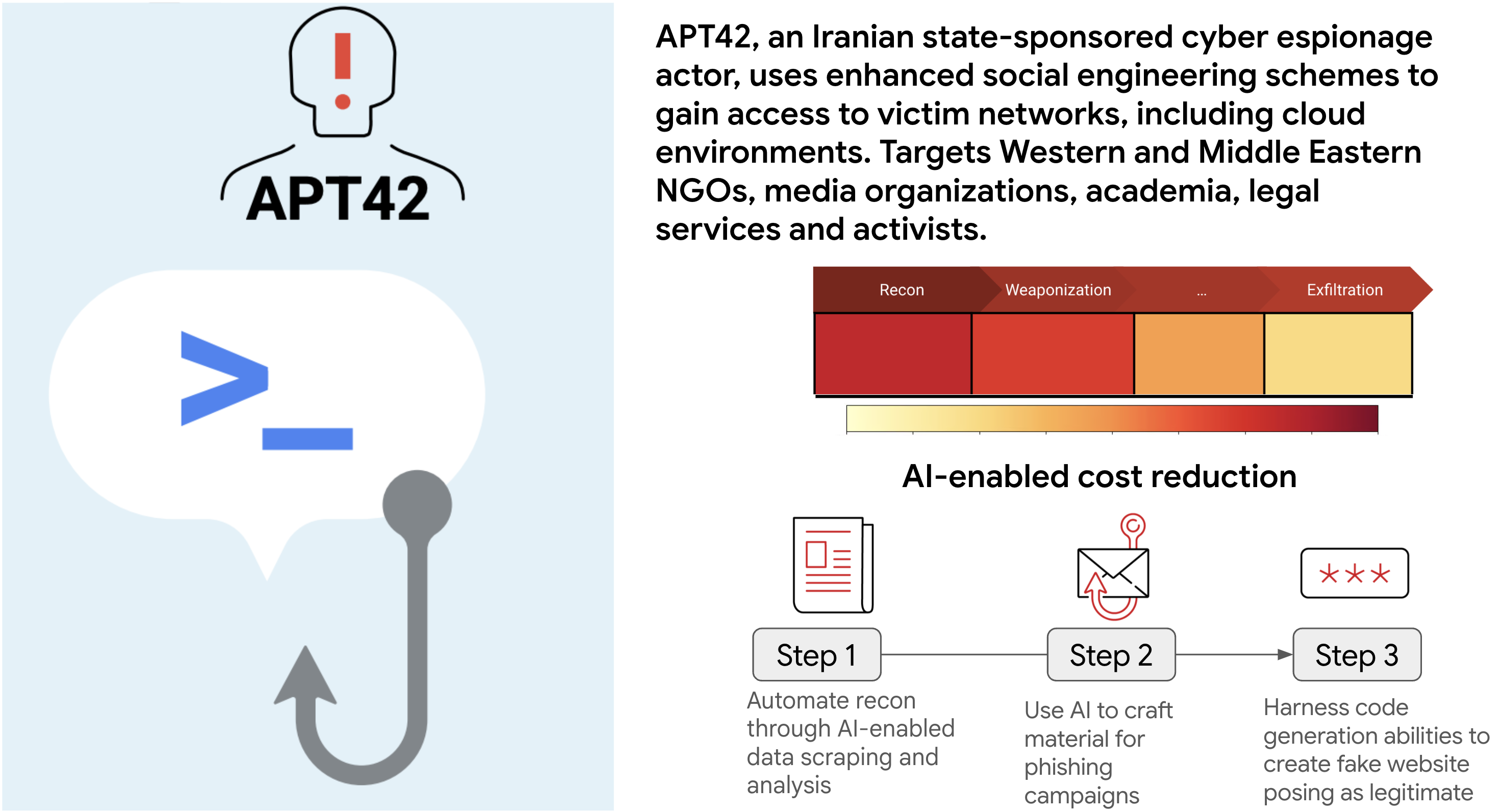}
  \caption{Framework enabling red teams to better model AI-enabled adversary behavior for testing defenses by generating emulation plans combining TTP knowledge with AI-enabled cost reduction evidence.}
  \label{fig:informing_red_teaming}
\end{figure*}

\textbf{Grounding AI-enabled Adversary Emulation.} The framework also informs proactive adversary emulation. Adversary emulation assesses security by applying threat intelligence about specific adversary TTPs to emulate threats, verifying detection/mitigation across the attack chain. Our framework helps red teams more accurately model AI-enabled adversary behavior (Figure \ref{fig:informing_red_teaming}). Combining knowledge of adversary TTPs, prevalence of AI use in specific phases (Figure \ref{fig:recon_ttps}), and evidence of AI-enabled cost reduction (Figure \ref{fig:heatmap_cost_reduction}) allows creation of more realistic emulation scenarios to test defenses against AI-leveraging actors.

\begin{figure*}[t] 
  \centering
  \includegraphics[width=\textwidth]{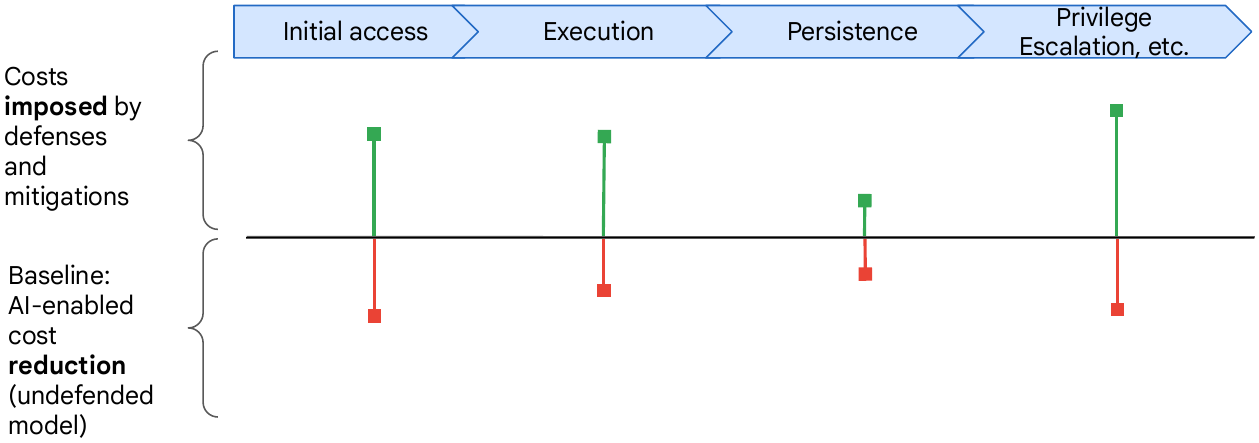}
  \caption{Benchmarking defensive intervention effectiveness by assessing cost imposition on AI-enabled attacks across the attack chain.}
  \label{fig:benchmarking_defenses}
\end{figure*}

\textbf{Benchmarking Defenses.} The approach can benchmark defense effectiveness by assessing costs imposed on leveraging AI for specific attack phases (Figure \ref{fig:benchmarking_defenses}). Cyber defense aims to increase attacker costs. While various defenses exist against AI-enabled attacks (model-level, post-deployment), a comprehensive framework for evaluating them across the attack chain is lacking. Our framework can assess intervention effectiveness in making AI-enabled attacks less efficient, potentially deterring them.

\section{Related Work}
AI has long been integral to cybersecurity, from malware detection to network analysis using predictive models. Recent frontier AI developments spurred research into defensive applications: vulnerability identification \citep{du2024vul,lu2024grace,li2021sysevr, akuthota2023vulnerability, al2024exploring}, incident summarization \citep{ban2023breaking,aminanto2020threat, khare2023understanding}, incident response \citep{hays2024employing}, and other foundational tasks \citep{alam2024ctibench}. DARPA’s AIxCC competition \citep{DARPA} showcased autonomous systems finding, exploiting, and fixing vulnerabilities \citep{ruan2024specrover,du2024vul,ristea2024ai}. However, the dual-use nature of cyber capabilities necessitates robust risk understanding and management. Consequently, research has grown on methods to evaluate potential risks of capable AI systems in cyber.

\subsection{Capture-the-Flag Challenges}
CTF challenges are the most common method for evaluating LLM offensive cyber capabilities. LLMs interface with CTF environments to solve security puzzles (cryptography, reverse engineering, web exploitation, etc.) by finding hidden 'flags'. Numerous works employ CTF benchmarks, including PentestGPT \citep{deng2023pentestgpt}, CyberSecEval 3 \citep{wan2024cyberseceval,bhatt2024cyberseceval}, Google DeepMind evaluations \citep{phuong2024evaluating}, PenHeal \citep{huang2023penheal}, AutoAttacker \citep{xu2024autoattacker}, Cybench \citep{zhang2024cybench}, EnIGMA \citep{abramovich2024enigma}, and InterCode-CTF \citep{yang2023intercode}. Some, like CyberSecEval 3, assess "copilot" scenarios with human operators using LLMs. Others focus on narrow benchmarks like Linux privilege escalation \citep{lu2024grace, ban2023breaking}, while CyberSecEval 3 covers a broader range but still limited attack phases. A drawback is the artificial constraints and simplified scenarios compared to real-world attacks (e.g., single target vs. complex enterprise networks), potentially skewing capability assessment.

\subsection{Multiple Choice and Free Response Tests}
Multiple-choice question benchmarks offer measurability and scalability \citep{wan2024cyberseceval, tihanyi2024cybermetric, liu2023secqa, tann2308using}. However, creating questions resistant to memorization that accurately reflect offensive cyber is challenging. CyberSecEval also uses free-response questions evaluated by another LLM. OCCULT \citep{kouremetis2025occult} introduced a multiple-choice benchmark for offensive tactic knowledge.

\subsection{Scaffolding and Capability Elicitation}
Research is emerging on capability elicitation and model scaffolding to measure upper-bound capabilities. Some systems are lightweight wrappers for action-observation loops (e.g., Cybench \citep{zhang2024cybench}, InterCode-CTF \citep{yang2023intercode}). Others offer moderate scaffolding (e.g., Vulnhuntr \citep{du2024vul}, AutoAttacker \citep{xu2024autoattacker}). More complex systems integrate extensive tools, multiple models, reasoning components, and human feedback (e.g., SWE Agent \citep{yang2024swe}, PentestGPT \citep{deng2023pentestgpt}, Project Naptime \citep{project_naptime}, EnIGMA \citep{abramovich2024enigma}, Incalmo \citep{singer2025feasibility}). While current evaluation approaches offer various tools, translating findings into actionable insights for defenders across the attack chain remains unclear. This paper aims to bridge this gap.

\section{Conclusion}
This paper introduced a novel framework for evaluating frontier AI's cyber capabilities, focusing on the end-to-end attack chain. Grounded in real-world AI misuse attempts, it bridges evaluations and defenses by helping prioritize targeted mitigations. We curated attack chain archetypes and a new benchmark, conducted bottleneck analysis to identify AI cost disruption potential, and showed how the framework illuminates cost impacts, facilitates mitigation prioritization, benchmarks defense effectiveness, and grounds AI-enabled adversary emulation.

Our evaluations revealed that current AI cyber assessments often overlook critical areas like evasion, detection avoidance, obfuscation, and persistence, where AI shows significant potential. We also confirm the importance of assessing misuse for reconnaissance, widespread exploitation, and long-term attacks.

Cybersecurity is dynamic, and AI will accelerate this. We expect AI to alter attack costs, prompting adversary adaptation. Our framework is designed to evolve with AI capabilities. We will continuously update attack chains, bottleneck analyses, and uplift evaluations based on real-world misuse and model evolution to provide defenders with decision-relevant insights. Mitigating misuse requires a community effort, including robust developer safeguards and evolving defensive techniques accounting for AI-driven TTP changes.

\section*{Acknowledgement}
We thank Lewis Ho for insightful reviews;  the larger Frontier Safety team for collaboration on frontier evaluations; Ivan Petrov for discussions on emerging capabilities; Jennifer Beroshi, Elie Bursztein, Xerxes Dotiwalla, Gena Gibson, Myriam Khan, Armin Senoner, Rohin Shah, Andy Song, and Andreas Terzis for guidance; Alexandru Totolici, Ansh Chandnani, Ash Fox, Daniel Fabian, Fu Chai, Maksim Shudrak, Mónica Carranza, Niru Ragupathy, Ryan Goosen, and Stefan Friedli, for their valuable time and insight into resource bottlenecks for historical cyberattack case studies; Pattern Labs specifically to Dan Lahav, Omer Nevo, Ofir Ohad, and Saar Tochner for access to their evaluation platform and library; Google's Threat Intelligence Group for relevant data and insights; and broadly Google for a supportive research environment.

\bibliography{main}
\appendix

\section{Bottleneck Analysis} \label{s:bottleneck}

\subsection{Initial analysis of bottlenecks}
We began the process of identifying an initial candidate list of bottlenecks by ingesting Mandiant's Threat Intelligence dataset. This data source consists of a set of detailed deconstructions of cyber attacks, timelines, and in-depth analysis of each phase of how an attack unfolded. The sources of this dataset include real-world data from Mandiant's work helping organizations recover from breaches, operational intelligence from monitoring and defending client networks, and adversarial intelligence obtained through in-depth research and analysis of threat actors.

\textbf{Phishing and spear phishing}

\textit{Bottlenecks: }
\begin{itemize}
\item Gathering information about the target (individual, organization, or group)
\item Creating a malicious payload (e.g., malware-laden attachment, malicious link) 
\item Transmitting the malicious payload to the target
\end{itemize}

\textit{AI Uplift Potential:}
\begin{itemize}
\item Automating Reconnaissance: AI-powered tools can quickly gather and analyze vast amounts of OSINT data to identify potential victims and tailor phishing lures.
\item Improving Weaponization: AI can generate highly convincing phishing emails and websites, making it more difficult for users to identify them as malicious.
\item Personalizing Attacks: AI can analyze individual user behavior to craft highly targeted phishing attacks that are more likely to be successful.
\end{itemize}

\textbf{Malware}

\textit{Bottlenecks:}
\begin{itemize}
\item Development of new malware: Requires reverse engineering, understanding vulnerabilities, coding skills, and potentially evading detection mechanisms.
\item Evading detection: Signature-based detection, sandboxing, behavioral analysis. Malware needs to be polymorphic or metamorphic to avoid detection.
\item Distribution/Delivery: Getting the malware onto target systems can be challenging. Exploiting vulnerabilities, social engineering, supply chain attacks.
\item Maintaining persistence: Ensuring the malware stays on the system even after reboots or security scans.
\item Command and Control (C2) communication: Establishing and maintaining covert communication channels for control and data exfiltration.
\end{itemize}

\textit{AI Uplift Potential:}

\begin{itemize}
\item Automated malware generation: AI can automate the creation of new malware variants, including polymorphic and metamorphic malware to evade signature-based detection.
\item Intelligent evasion techniques: AI can learn and develop techniques to evade sandboxing and behavioral analysis by mimicking benign behavior or detecting sandbox environments.
\item Automated vulnerability exploitation: AI can be used to find and exploit vulnerabilities to deliver and install malware automatically.
\item Enhanced C2 communication: AI can establish more resilient and stealthy C2 channels, potentially using techniques like domain generation algorithms (DGAs) or encrypted communications that adapt to network conditions.
\item Targeted malware: AI can tailor malware payloads and behaviors to specific targets, increasing effectiveness and reducing detection.
\end{itemize}

\textbf{Denial-of-Service (DoS)}

\textit{Bottlenecks:}
\begin{itemize}
\item Amplification: Generating enough traffic to overwhelm a target infrastructure can be difficult without amplification techniques.
\item Bypassing mitigation strategies: Rate limiting, firewalls, intrusion detection/prevention systems, content delivery networks (CDNs).
\item Maintaining attack persistence: Keeping the attack going continuously can be resource intensive, and mitigation strategies might eventually become effective.
\item Attribution and anonymity: Hiding the source of the attack can be challenging and important for avoiding repercussions.
\end{itemize}

\textit{AI Uplift Potential:}

\begin{itemize}
\item Intelligent amplification attacks: AI could optimize amplification techniques to maximize the impact of DoS attacks with fewer resources, potentially by dynamically adapting attack vectors.
\item Automated DDoS orchestration: AI can automate the orchestration of large-scale DDoS attacks, managing botnets and attack vectors more efficiently.
\item Evasion of mitigation: AI can learn and adapt to bypass rate limiting, firewalls, and other mitigation strategies by identifying weaknesses in defensive systems and dynamically changing attack patterns.
\item Creation of more complex and stealthy DoS attacks: AI might enable development of application-layer DoS attacks that are harder to detect and mitigate than simple volumetric attacks.
\item Autonomous botnet management: AI could manage botnets more autonomously and effectively, improving their resilience and attack capabilities.
\end{itemize}

\textbf{Man-in-the-Middle (MitM)}

\textit{Bottlenecks:}

\begin{itemize}
\item Network positioning: Gaining a position on the network path between two communicating parties (e.g., ARP poisoning, rogue Wi-Fi access points).
\item Traffic interception: Capturing and potentially decrypting network traffic. Encryption (HTTPS, TLS) makes interception and decryption harder.
\item Real-time traffic analysis: Analyzing intercepted traffic in real-time to extract valuable information or identify opportunities for manipulation.
\item Traffic manipulation/injection: Modifying traffic without being detected, which requires understanding the protocols and application logic.
\item Maintaining stealth: Avoiding detection while intercepting and potentially manipulating traffic.
\end{itemize}

\textit{AI Uplift Potential:}

\begin{itemize}
\item Automated network positioning: AI can automate network reconnaissance and identify optimal positions for MitM attacks.
\item Intelligent traffic analysis: AI can perform deep packet inspection and real-time analysis of encrypted traffic to identify patterns, vulnerabilities, or sensitive data even without full decryption, potentially using techniques like traffic analysis and machine learning.
\item Dynamic traffic manipulation: AI could automate the dynamic manipulation of traffic based on real-time analysis, enabling more sophisticated and context-aware attacks.
\item Bypassing encryption or finding weaknesses in implementations: AI could potentially find subtle weaknesses in encryption protocols or implementations that can be exploited for partial or full decryption in certain scenarios.
\item Automated injection of malicious content: AI can inject malicious content into traffic streams in a way that is less likely to be detected and more likely to achieve the attacker's objectives
\end{itemize}
.
\textbf{SQL Injection}

\textit{Bottlenecks:}

\begin{itemize}
\item Finding vulnerable parameters: Identifying input fields in web applications that are vulnerable to SQL injection.
\item Crafting effective injection payloads: Developing SQL queries that can bypass input validation and achieve the desired outcome (data exfiltration, modification, etc.).
\item Bypassing web application firewalls (WAFs): WAFs are designed to detect and block common SQL injection attacks.
\item Exploiting complex SQL injection scenarios: Blind SQL injection, time-based injection, second-order injection can be more complex to exploit.
\item Automating the exploitation process: Manually testing for and exploiting SQL injection can be time-consuming.
\end{itemize}

\textit{AI Uplift Potential:}

\begin{itemize}
\item Automated vulnerability scanning and identification: AI can crawl web applications and automatically identify potential SQL injection vulnerabilities with greater accuracy and speed.
\item Intelligent payload crafting: AI can generate SQL injection payloads that are more likely to bypass input validation and WAFs, potentially using techniques like mutation and adversarial examples.
\item Automated exploitation of complex scenarios: AI can automate the exploitation of blind, time-based, and second-order SQL injection vulnerabilities, significantly reducing the time and effort required.
\item Learning WAF evasion techniques: AI can learn from WAF responses and develop evasion techniques that are more effective.
\item Optimized data exfiltration: AI can optimize data exfiltration strategies after successful SQL injection to minimize detection and maximize data retrieved.
\end{itemize}

\textbf{Zero-Day Attack}

\textit{Bottlenecks:}

\begin{itemize}
\item Vulnerability discovery: Finding previously unknown vulnerabilities is extremely difficult and time-consuming, requiring deep expertise and resources.
\item Exploit development: Creating a reliable exploit for a zero-day vulnerability that works across different systems and is not easily detected.
\item Weaponization and delivery before patching: Attacks need to be carried out before the vulnerability is publicly disclosed and patched, requiring speed and stealth.
\item Maintaining secrecy of the vulnerability: Keeping the zero-day vulnerability secret is crucial for its long-term effectiveness.
\item Target selection and impact maximization: Choosing targets where the zero-day exploit will have maximum impact.
\end{itemize}

\textit{AI Uplift Potential for Zero-Day Attacks:}

\begin{itemize}
\item Accelerated vulnerability discovery: AI can analyze codebases and software systems at scale to identify potential zero-day vulnerabilities much faster than traditional methods, using techniques like fuzzing, symbolic execution, and machine learning for anomaly detection in code.
\item Automated exploit generation: AI can automate the process of generating exploits for discovered vulnerabilities, reducing the time and attack barrier for exploit development.
\item Proactive vulnerability prediction: AI might be able to predict potential vulnerability types or locations in software based on code patterns and past vulnerability data, guiding vulnerability research efforts.
\item Stealthy zero-day weaponization: AI can help create zero-day exploits and delivery mechanisms that are more stealthy and harder to detect, maximizing the window of opportunity before patching.
\item Targeted zero-day attacks: AI can analyze potential targets and identify those where a specific zero-day exploit
\end{itemize}

\subsection{Validating Attack Bottlenecks via Expert Survey}

\begin{figure*} 
\centering
\includegraphics[width=\textwidth]{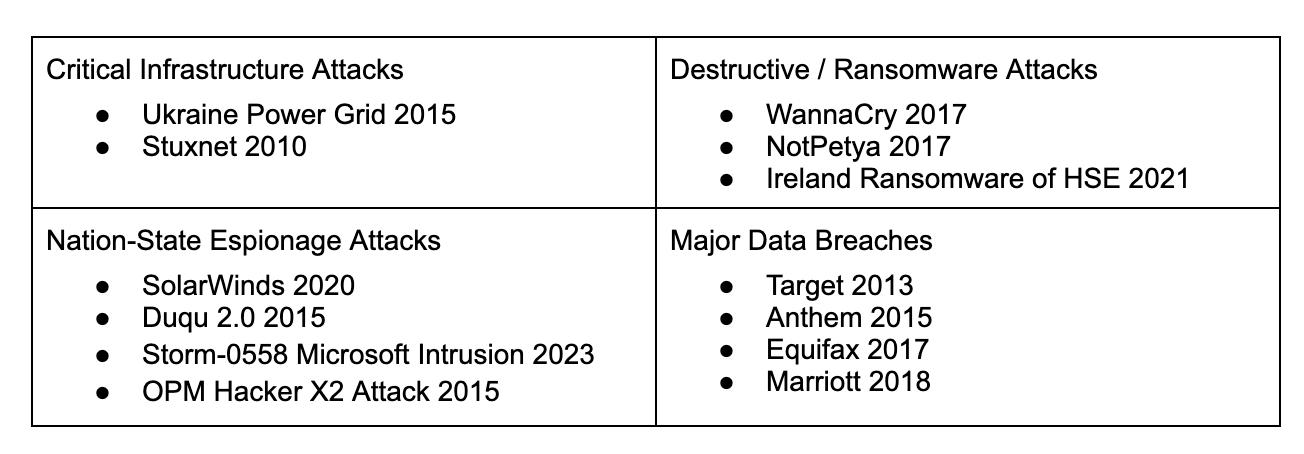} 
\caption{Notable, high-impact case studies selected for expert verification.}
\label{fig:case_studies}
\end{figure*}

To validate potential bottlenecks in cyberattacks, we first selected thirteen historical case studies exhibiting exceptionally high impact from a broader set of representative attack chains. The selection criteria included attacks that caused, or had the potential to cause, catastrophic loss of life, significant economic damage, or state-level cyber espionage. (See Figure \ref{fig:case_studies}). Assessing the economic impact of attacks is inherently difficult; company reports might overestimate losses (as revenue may shift rather than disappear), while negative supply chain effects can be underestimated. Therefore, impact assessment requires considerable subjective judgment.

We identified suitable case studies using sources such as the Kent Academic Repository \citep{johansmeyer2024perception}, CSIS reports, Florian Roth's summaries, security company publications, and media reports. To focus on the modern defensive landscape, we excluded attacks that occurred before 2010.

Next, we recruited ten offensive security experts from Google. They estimated the resources required to replicate the selected attacks under specific assumptions: the attack would be performed by a hypothetical median-skilled cyber expert at Google, without using AI tools. This standardization helps ensure comparable estimates regardless of the original attacker's capabilities.

The survey proceeded as follows:

\begin{enumerate}
\item Each expert estimated the resources needed for each phase of two distinct case studies. We provided detailed descriptions of the attacker's actions and achieved outcomes for every phase.
\item Experts provided estimates in two forms: human-days of effort for a median expert, and direct monetary expenses (e.g., hardware procurement, infrastructure costs). They also rated their confidence (low, medium, high) for each estimate. Following \citet{haga2020modelling}, we broke down attacks into phases to simplify estimation and requested cost intervals rather than point estimates to improve generalizability.
\item To establish a plausible upper bound on phase efficiency and avoid assuming replication of historical attacker errors, experts also estimated the minimum resources each phase might require for a sophisticated adversary.
\item This process resulted in seven case studies reviewed by two experts and six reviewed by one. We then conducted a consensus-building exercise: for doubly-reviewed cases, experts analyzed their peer's estimates and reasoning and could optionally revise their own; for singly-reviewed cases, a second expert provided anonymous feedback on the initial assessment.
\end{enumerate}

This survey methodology helps determine the relative costs of different attack phases and identify dominant bottlenecks. Understanding these bottlenecks allows defenders to qualitatively assess how enhancing defenses at one stage might affect the overall attack economics and potentially anticipate significant shifts in attack costs.

We took the initial human-day estimates and converted them to dollar costs by multiplying by a salary variable (\$500k); for ease of cost comparison between phases, we then calculated the percent of the overall resourcing estimate each phase accounted for. We define a 'bottleneck' as any phase requiring at least 10\% of the total estimated resources for the attack.

Our key findings were that:
\begin{itemize}
    \item {Weaponization and Reconnaissance were very common bottlenecks, and also commanded the overall highest amount of resourcing (averaged across scenarios where they were bottlenecks)}
    \item {For major data breaches, bottlenecks  were  more  evenly  distributed across phases compared to other attack types}
    \item {Averages across case studies:
        \begin{itemize}
            \item Weaponization (n=9, 35.94\%, \$206.7k)
            \item Reconnaissance (n=7, 19.20\%, \$44.8k)
            \item Actions on Objectives (n=5, 17.60\%, \$525k; 11.00\%, \$16.5k when excluding SolarWinds outlier)
            \item Exploitation  (n=4, 9.82\%, \$19.3k)
            \item Delivery (n=4, 8.30\%, \$15.8k)
            \item C2 (n=3, 6.40\%, \$17.3k)
            \item Installation (n=2, 3.20\%, \$8.3k)
        \end{itemize}
    }
\end{itemize}
\end{document}